\documentclass[preprint,showpacs,preprintnumbers,amsmath,amssymb]{revtex4}

\usepackage{graphicx}% Include figure files
\usepackage{dcolumn}% Align table columns on decimal point
\usepackage{bm}% bold math
\usepackage{color}
\usepackage{mathrsfs}

\newcommand{\F}{{\mathscr F}}

\begin{document}

\preprint{APS/123-QED}

\title{Characterization of intermittency in renewal processes:\\ Application to earthquakes}% Force line breaks with \\

\author{Takuma Akimoto}
\email{akimoto@aoni.waseda.jp}
% \altaffiliation[Also at ]{Physics Department, Waseda University.}%Lines break automatically or can be forced with \\
\author{Tomohiro Hasumi}%
\author{Yoji Aizawa}
 
\affiliation{%
Department of Applied Physics, Advanced School of Science and Engineering, Waseda University, Okubo 3-4-1, Shinjuku-ku, Tokyo 169-8555, Japan.
%\\ This line break forced with \textbackslash\textbackslash
}%

%\affiliation{

\date{\today}% It is always \today, today,

\begin{abstract}
We construct a one-dimensional piecewise linear intermittent map from the interevent  
 time distribution  for a given renewal process. 
Then, we characterize intermittency by the asymptotic behavior near the indifferent fixed point in the piecewise linear intermittent map. Thus, we provide a new framework 
to understand a unified characterization of intermittency, and also present 
the Lyapunov exponent of renewal processes.
This  method is applied to the occurrence of 
earthquakes using the Japan Meteorological Agency (JMA) catalog. We demonstrate that interevent times are not independent and identically distributed random variables by analyzing  the return map of interevent times, but that there is a systematic change in conditional probability  distribution functions of interevent times. 
\end{abstract}

\maketitle

\section{\label{sec:level1} Introduction}

Recently, intermittent phenomena, characterized by a power law of 
the laminar state, have attracted interest in non-equilibrium statistical 
physics as well as biological and atomic physics. Examples of 
intermittent phenomena are the fluorescence of quantum dots 
\cite{Haase2004} and nanocrystals \cite{Brok2003}, and ion channel 
gating \cite{Lowen1999}. Earthquakes can be recognized as an 
intermittent phenomenon. Actually, a $1/f^{\delta}$ spectrum is observed in 
the $P$-wave and $S$-wave velocity \cite{Shiomi1997}. Also, the residence time distribution of the laminar state, in which the number of earthquakes per 
unit time is lower than a threshold, obeys a power law \cite{Bottiglieri2007}, 
and the intermittency for the occurrence of earthquakes in Irpinia, Italy, has been  
quantified using the correlation codimension \cite{Telesca2001}. 
Moreover,  intermittency appears in the stick-slip model of earthquakes 
\cite{Ryabov2001}. In such a non-Poisson processes, a
long time tail and aging have been clearly observed \cite{Allegrini2004, Margolin2006}. Non-hyperbolic dynamical systems, which have at least one indifferent 
fixed point, also show intermittent behavior such as a long time tail and 
non-stationarity \cite{Ai1984, Akimoto2007}. By applying renewal theory to 
symbolic dynamics generated by the coarse graining of an orbit, it was also 
shown that non-hyperbolic dynamical systems generate a $1/f$ spectrum 
\cite{Ai1984, Ben-Mizrachi1985}.\par
An outstanding problem in intermittent phenomena is the  
incompleteness of the usual statistical descriptors such as mean and variance 
because of the divergence of the mean interevent time \cite{FN2}.
 It 
is  remarkable that infinite measure preserving dynamical systems, which 
are closely related to intermittent phenomena, exhibit intrinsic 
non-stationarity \cite{Akimoto2008}. Even when the mean interevent time 
is finite, the long tail of distribution makes it difficult to characterize 
the interevent time by the mean. Usually,  intermittency is characterized by the exponent of a power law. However, this characterization does not 
include a long tail distribution heavier or not heavier than a power law. 
In the present paper, we characterize intermittency to estimate the difficulty of forecasting rare events. Moreover, the degree of activity of events is studied from the viewpoint of a non-hyperbolic dynamical system.
\par
Renewal processes have been drawing attention not only in mathematics 
but also in physics, and are useful to analyze intermittent phenomena \cite{God2001}. In fact,
intermittent phenomena can be rewritten as renewal processes 
by focusing  attention on the residence time distribution of laminar state. 
In renewal processes, it is 
assumed that the interevent times between renewals are independent and 
identically distributed (i.i.d.) random variables. 
For a given renewal process we construct  a one-dimensional dynamical system to characterize intermittency in renewal processes. Then, we develop a concept of intermittency based on  
dynamical systems.  One of our results is a unified  characterization of intermittency 
in renewal processes, where we classify renewal processes into 
six different regimes according to difficulty to forecast the next event: (i) non-stationary essential singular intermittency, 
(ii)  non-stationary very strong intermittency, 
(iii) stationary strong intermittency, (iv)  stationary weak intermittency, 
(v)  stationary very weak intermittency, and (vi)  stationary non-intermittent 
chaos.\par

The paper is organized as follows. In Sec. II, we construct one-dimensional piecewise linear intermittent maps  from renewal processes with any distribution, and then intermittency is characterized by the asymptotic behavior near the indifferent fixed point. Using the constructed map, we can calculate the Lyapunov exponent of a renewal process. In Sec. III, we study the occurrence of earthquakes. To apply our method to the occurrence of earthquakes in Japan, we verify whether the occurrence of earthquakes is a renewal process or not.  Then, our method is applied to  the occurrence of earthquakes by analyzing the conditional probability distribution functions of interevent times.
 Conclusions are given at the end of the paper.

%\subsection{\label{sec:level2}Second-level heading: Formatting}

%\subsubsection{\label{sec:level3}Third-level heading: References and Footnotes}

%\begin{figure}
%\includegraphics[height=.5\linewidth]{fig3.eps}% Here is how to import EPS art
%\caption{Sequence of earthquakes with magnitude $M\geq 2$ for JMA catalog from 
%January 12, 2001 to January 14, 2001. Three time intervals between renewals 
%are indicated by arrows, where renewals are defined as the occurrence of an 
%earthquakes with magnitude $M\geq 3.9$.}
%\end{figure}

\section{Underlying dynamical systems in renewal processes}

\subsection{Construction of one-dimensional maps}

To characterize intermittency, we construct  one-dimensional 
maps from discrete time renewal processes. Let $f(m)$  be 
the probability distribution function of a random variable $m$ $(m=1,2,\cdots)$, and $F(m)=\sum_{k=1}^m f(k)$
and $\F (m)=1-F(m)$, where $F(0)=0$ and $\F(0)=1$. Then, we can obtain the following relationship: 
\begin{equation}
f(m) = \F (m-1) -\F(m).
\label{eq:2.1}
\end{equation}
Using this relation, we can construct a one-dimensional map on $[0,c]$ 
in which the residence times in $[0,1]$ are i.i.d. random variables with 
probability $f(m)$. Concretely, the map is given by a piecewise linear map $T : [0,c] \rightarrow [0,c]$ 
defined by 
\begin{eqnarray}
x_{n+1}&=&T(x_n)\nonumber\\
&=&\left\{
\begin{array}{ll}
{\displaystyle
\frac{a_{k-1}-a_k}{a_k-a_{k+1}}(x_n-a_k)+a_{k-1}},
\quad &x_n\in [a_{k+1},a_{k}),\\
\\
{\displaystyle
\frac{x_n-1}{c-1}}, &x_n\in [1,c],
\end{array}
\right.
\label{eq:2.2}
\end{eqnarray}
where a sequence $a_k$ is given by $a_k=\F(k)$ 
and $\F(-1)=c$. Actually, a point in the 
interval $[a_m,a_{m-1}]$ is mapped into the interval $[1,c]$ by $m+1$th
 iterations, and then the probability of the residence time $m$ 
in the interval $[0,1]$ is given by $a_m-a_{m-1}$ (see Fig. 1).\par

\begin{figure}
\includegraphics[height=5cm,width=5cm, angle=0]{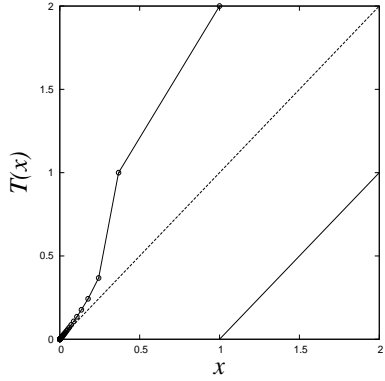}% Here is how to import EPS art
\caption{Piecewise linear map $T(x)$ for the Weibull distribution $(a=0.5, \tau=5, c=2.)$. Circles indicate the end points of the straight line segments.}
\end{figure}

Next, we focus on the constructed map near the fixed point $(x=0)$. 
The derivative of the map is given by 
\begin{equation}
T'(x)|_{x\in [a_n,a_{n-1})}=\frac{a_{n-1}-a_{n}}{a_n-a_{n+1}}
=\frac{f(n)}{f(n+1)}.
\label{eq:2.3}
\end{equation}
Considering the maps near the fixed point, i.e., $x\cong a_n=\F (n)$, 
$a_{n-1}-a_n\cong 0$, we obtain the 
asymptotic form, 
\begin{equation}
T'(x)\sim \frac{f(\F^{-1}(x))}{f(\F^{-1}(x)+1)} \quad~{\rm as}~
x\rightarrow 0.
\label{eq:2.4}
\end{equation}
In particular, the asymptotic behavior for a power law distribution 
$(\F (m) \sim m^{-\beta})$ is given by
\begin{equation}
T'(x) -1 \propto x^{1/\beta} \quad {\rm as}~
x\rightarrow 0.
\label{eq:2.5}
\end{equation}
This map is the same as the Pomeau-Manneville map \cite{PM1979}, which 
is a typical example of intermittent maps, and the piecewise linear 
version is also well studied as an intermittent map \cite{Tasaki2002, Miyaguchi2007}. The asymptotic behavior is given by 
\begin{equation}
T'(x) \sim e^{1/\tau} \quad {\rm as}~
x\rightarrow 0~{\rm for}~\F(m) \sim e^{-m/\tau},
\label{eq:2.6}
\end{equation}
and
\begin{equation}
T'(x)-1 \propto (-\log x)^{(a-1)/a} \quad {\rm as}~
x\rightarrow 0
\label{eq:2.7}
\end{equation}
for the Weibull distribution $(\F(m)\sim e^{-(m/\tau)^a})$ $(a\ne 1)$.
In the case of the log-Weibull distribution $(\F(m) \sim e^{-(\log m/\tau)^b})$, 
the asymptotic behavior is given by 
\begin{equation}
T'(x)-1 \propto \exp( -\tau(-\log x)^{1/b}) \quad {\rm as}~
x\rightarrow 0. 
\label{eq:2.8}
\end{equation}
We refer to Eqs. (7) %\ref{eq:2.7})
 and (8)
% (\ref{eq:2.8})
 as the Weibull map 
and the log-Weibull map, respectively. 
 Note that the Weibull map with exponent $a<1$ and the log-Weibull 
map have the indifferent fixed point $(T'(0)=1)$. \par

\begin{figure}
\begin{minipage}{.5\textwidth}
\includegraphics[height=.33\linewidth,angle=0]{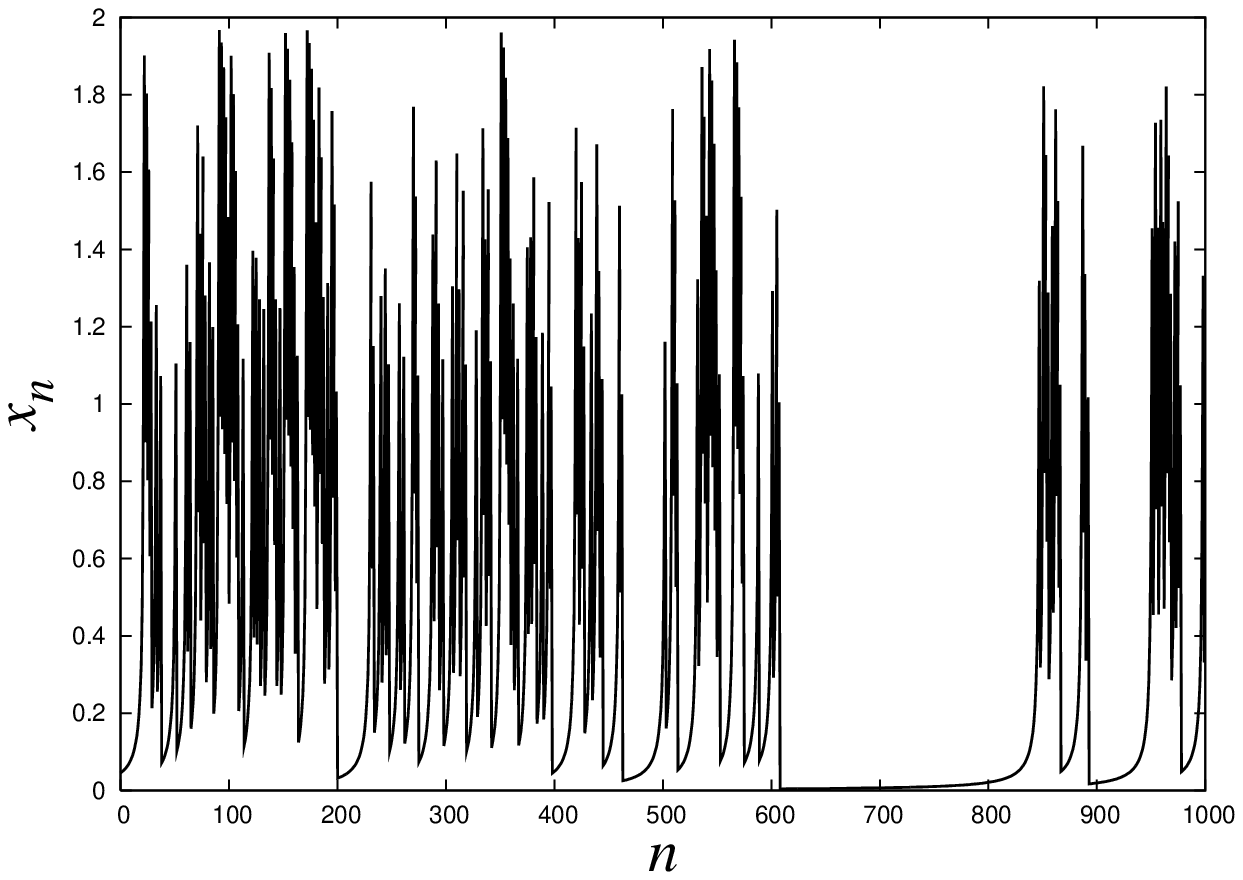}
\includegraphics[height=.33\linewidth,angle=0]{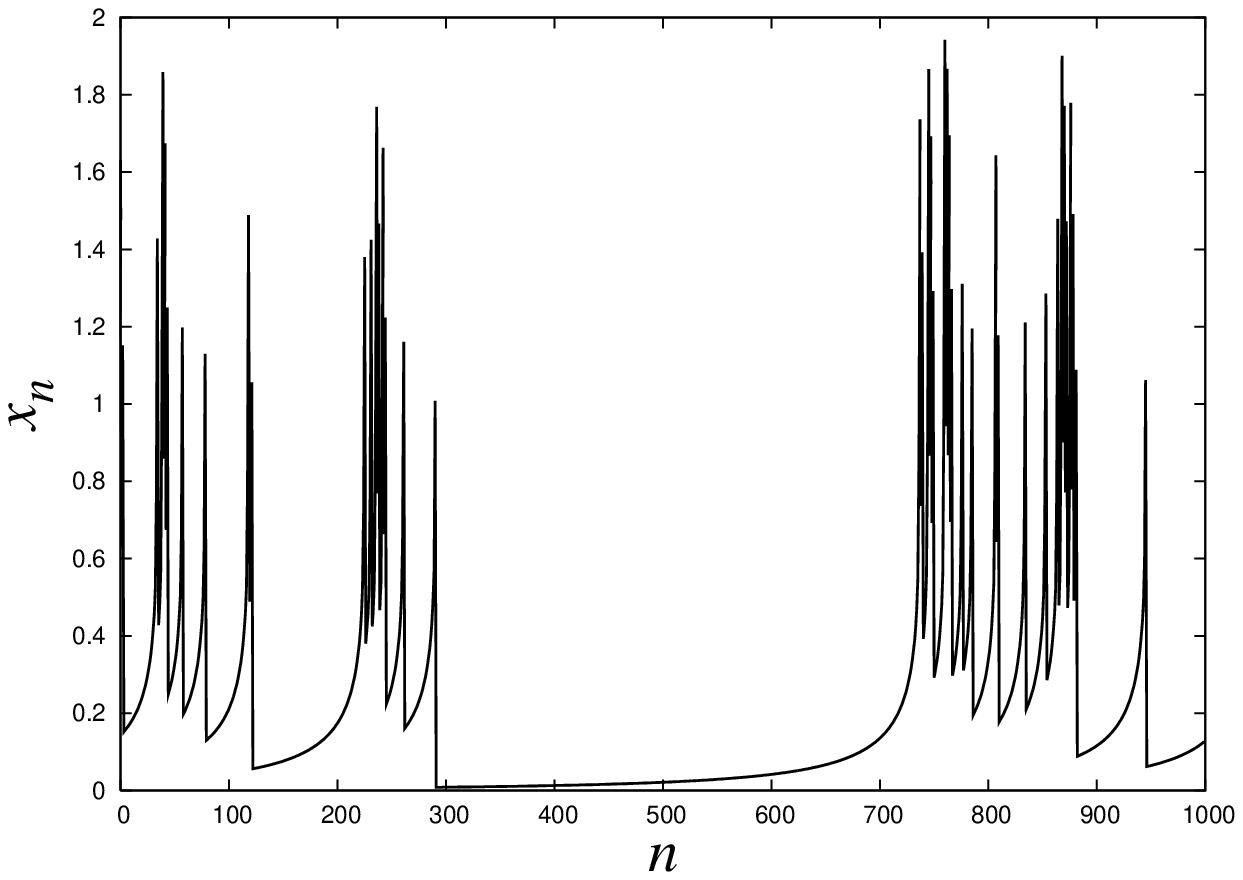}
\end{minipage}
\caption{Time series of piecewise linear intermittent maps. The map in the left figure is constructed using the power law distribution $(\F (m)=m^{-1})$. The map in the right figure is constructed using the Weibull distribution $(\F (m) = \exp(-(m/5)^{0.35})$.}
\end{figure}

\subsection{Characterization of  intermittency}

In intermittent chaos, an orbit stagnates near indifferent fixed points for 
an extremely long time (laminar state), and then irregular chaotic motion occurs (see Fig. 2). The residence time distribution of the laminar state is determined by 
the structure of a map near the indifferent fixed points. Here, we 
characterize intermittency from the asymptotic behavior of the derivative at 
the indifferent fixed point $(x=0)$:
\begin{equation}
T'(x)-1 \propto A(x)x^{\alpha} \quad {\rm as}~
x\rightarrow 0,
\label{eq:3.1}
\end{equation}
where $T(x)$ is the one-dimensional map constructed by a renewal process and 
$A(x)=o(1)$ and $x^{\alpha}=o(A(x))$. The degree of intermittency is 
classified into six types:
\begin{enumerate}
\item[(i)] $\alpha=\infty$, i.e., $T'(x)-1 \propto e^{-x^{-\alpha'}}$ as $x\rightarrow 0$ $(\alpha'>0)$:  non-stationary essential singular intermittency,
\item[(ii)] $\alpha\geq 1$ :  non-stationary very strong intermittency,
\item[(iii)] $0<\alpha<1$ :  stationary strong intermittency,
\item[(iv)] $\alpha=0$ and $A(x) =o(L(x))$ as $x\rightarrow 0$ :  stationary weak intermittency,
\item[(v)] $\alpha=0$ and $A(x) \sim L(x)$ as $x\rightarrow 0$ :  stationary very weak intermittency,
\item[(vi)] $T'(x)-1 >0$ as $x\rightarrow 0$ : stationary non-intermittent
 chaos,
\end{enumerate}
where the function $L(x)$ is slowly varying at 0 \cite{FN3}. The intensity of  intermittency can be quantified by the exponent $\alpha$. The larger $\alpha$ is, the more difficult to forecast  the next event  becomes. This is because slightly different reinjections near the fixed point make the residence time in $[0,1]$ completely different in the case of large $\alpha$.
With regard to type (iv), to be more precise, 
\begin{equation}
T'(x)-1 = O(\exp (-(-\log x)^{\gamma}))\quad {\rm as}~
x\rightarrow 0,
\label{eq:3.2}
\end{equation}
we can quantify the intensity of intermittency by the exponent $\gamma$. 
For type (v), 
\begin{equation}
T'(x)-1 = O\left(\frac{1}{ (-\log x)^{\eta}}\right)
 \quad {\rm as}~
x\rightarrow 0, 
\label{eq:3.3}
\end{equation}
so that the intensity of intermittency is determined by the exponent $\eta$.  
Exponents $\alpha,\gamma$ and $\eta$ represent the degree of the difficulty to forecast events in renewal processes because the sensitive dependence of the interevent time on reinjection points is determined by these exponents.\par
Note that the renewal function does not increase linearly 
with time; that is, the occurrence of renewals is not stationary, when the mean 
of the interevent times is not finite \cite{Cox}. Accordingly, 
the occurrence of renewals becomes non-stationary in the case 
of $\gamma\geq 1$. For $\gamma>1$,  intermittency is classified 
into the non-stationary essential singular intermittent regime.
\par

\subsection{ Lyapunov exponent}
We can calculate the Lyapunov exponent in renewal processes because we constructed one-dimensional maps from renewal processes. In general, 
the Lyapunov exponent $\lambda$ is defined by 
\begin{equation}
\lambda =\lim_{n\rightarrow \infty}\frac{1}{n}\sum_{k=1}^n 
\ln |T'(x_k)|,
\label{eq:2.3.1}
\end{equation}
where $T(x)$ is a one-dimensional map constructed from a renewal process. 
The slope of the piecewise linear map on $[a_1,a_0)$ and $[1,c]$ is 
given by $(c-1)/(1-a_1)$ and $1/(c-1)$, respectively. Therefore, 
the Lyapunov exponent does not depend on $c$ because $\ln (c-1)/(1-a_1)+
\ln 1/(c-1)=-\ln (1-a_1)$.
However, the Lyapunov exponent strongly depends upon the unit of 
time in renewal processes. Physical meaning of the Lyapunov exponent is the degree of activity of events. In other words, a large Lyapunov exponent implies  high activity of an underlying dynamical system.

\section{Application to the occurrence of earthquakes}

We apply this method to the occurrence of earthquakes using 
the JMA catalog \cite{FN4}
 in the area enclosed within 25${}^{\circ}$-50${}^{\circ}$N latitude and 
125${}^{\circ}$-150${}^{\circ}$E longitude with magnitude $M\geq 2$. 
We are interested in the interevent time distribution for a tail 
part to construct a one-dimensional piecewise linear map. Similar 
to previous studies \cite{Corral2004, Hasumi2009}, we consider 
 earthquakes with magnitude above a certain threshold $M_c$.
In other words, we study the interevent time statistics in which 
magnitude is greater than $M_c$. Here, we use the earthquake data from 
January 1, 2001 to October 31, 2007. 
\par
To verify the hypothesis that the occurrence of earthquakes is a renewal process, we analyze the return map of interevent times. As shown in Fig 3, the average of an interevent time will be small if the previous interevent time is small. Thus, an interevent time depends clearly on the one preceding it.  However, the average of an interevent time is constant when the previous one is relatively large, which suggests that interevent times are i.i.d. random variables.\par 
To analyze the dependence of the distribution of interevent times on the previous one in detail,  we sorted data in order and divided it into ordered data sets. To avoid statistical error, the number of data in each ordered data set is fixed at $10^4$.
Analyzing the conditional probability distribution functions of interevent times for each ordered data set, we find that they change systematically. Moreover, we find that the conditional probability distribution functions converge to the {\it Weibull distribution} according to increases of  previous interevent times, where  the Weibull distribution 
$F(t)$ is defined by 
\begin{equation}
F(t) =1- \exp (-(t/\tau)^a), 
\label{eq:4.1}
\end{equation}
(see Figs. 4 and 5). It is  remarkable  that  all conditional probability distribution functions obey the {\it Weibull distribution} in a tail region. 
Therefore,  the intermittency of the occurrence of earthquakes is {\it stationary very weak intermittency} because the conditional probability distribution functions of interevent times in the tail region are invariant and described by the Weibull distribution.
\par
Analyzing the distribution  for different $M_c$, 
we find that the conditional probability distribution functions  obey the Weibull distribution 
when the previous interevent time is relatively large. The Weibull exponent $a$, the intensity of intermittency $\eta$, and the Lyapunov exponent $\lambda$   are summarized in Table I, 
where the time unit is set to be one second in the calculation of
the Lyapunov exponent.

\begin{figure}
\includegraphics[height=1.\linewidth, angle=-90]{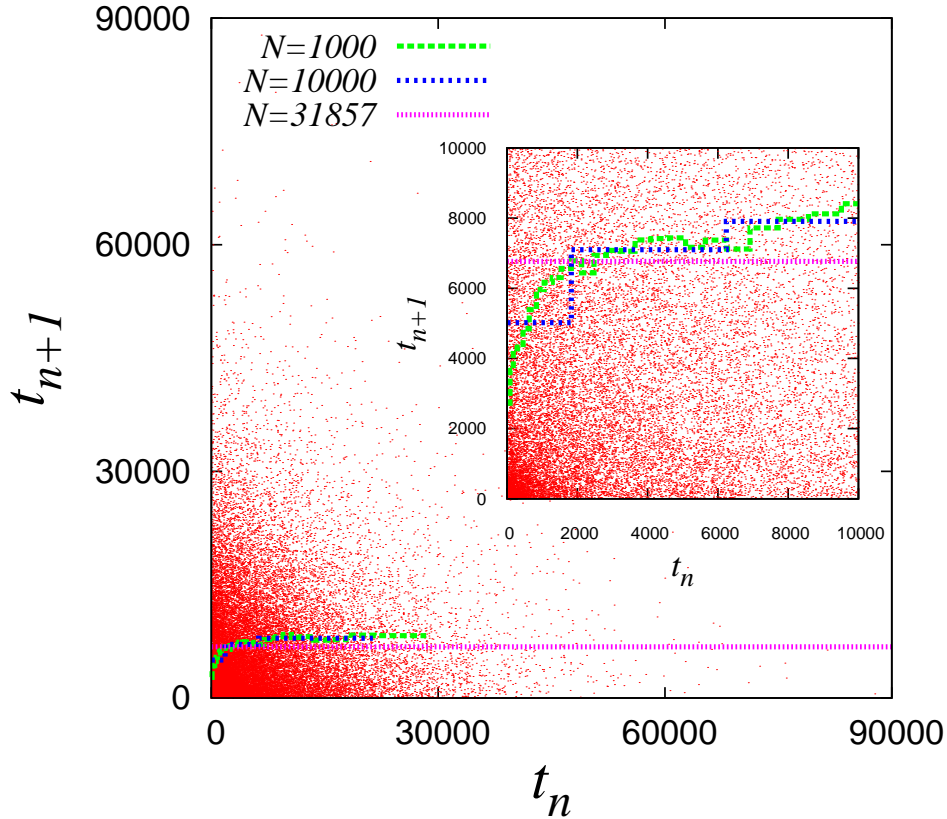}% Here is how to import EPS art
\caption{(color online). Return map of interevent times in Japan ($M_c=3.0$). The dotted lines are averages of the interevent time, where averages are taken in three different bins $(N=1000, 10000$ and $31857)$. Inset figure is a blow-up of return map ($M_c=3.0$). }
\end{figure}

%\begin{figure}
%\includegraphics[height=.9\linewidth, angle=-90]{dist.RM.mc=2.0.sort.eps}% Here is how to import EPS art
%\caption{Return map ($M_c=2.0$). }
%\end{figure}

\begin{figure}
\includegraphics[height=.9\linewidth, angle=-90]{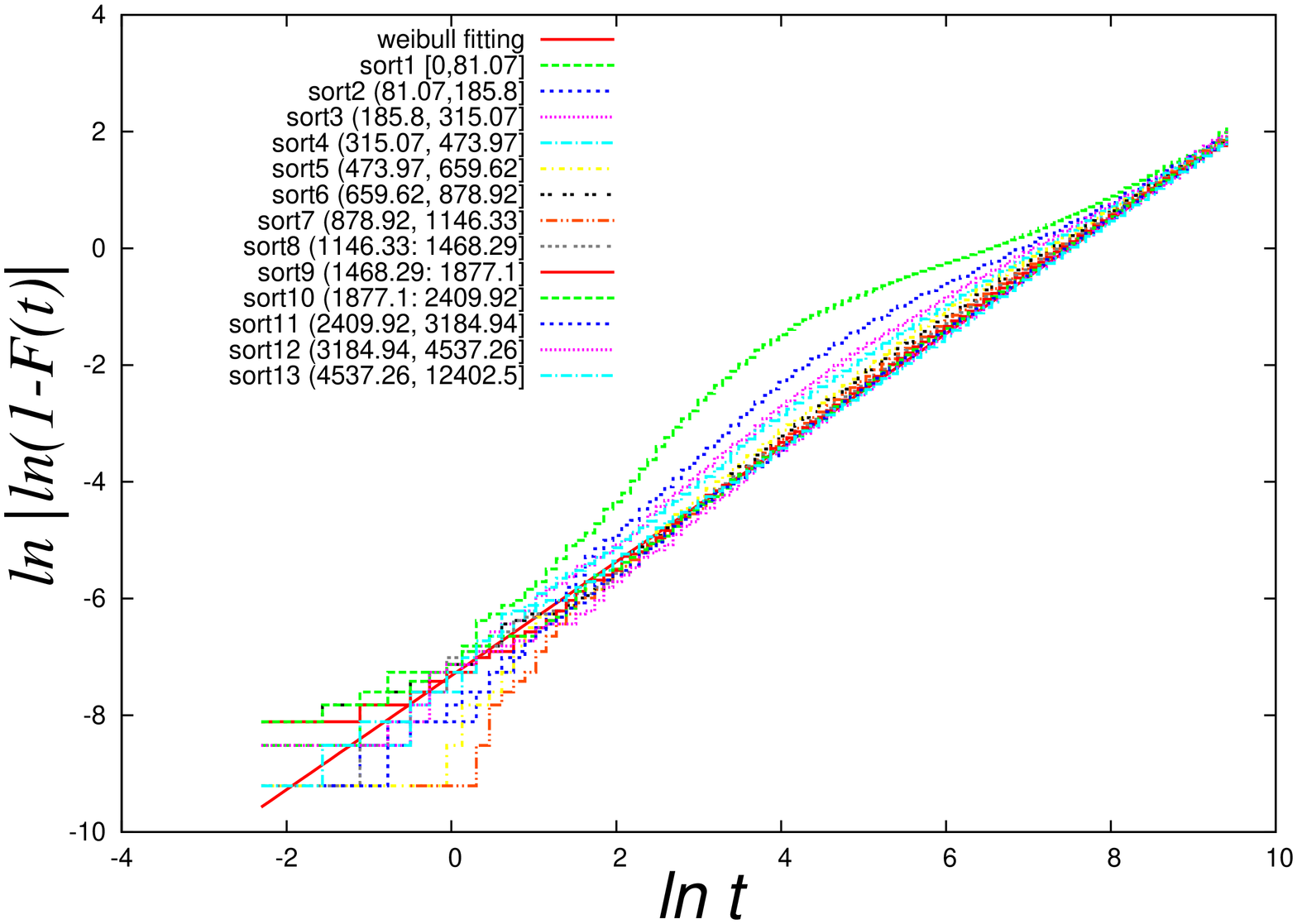}% Here is how to import EPS art
\caption{(color online). Weibull plot  of the distribution of 
 interevent times of earthquakes in Japan $(M_c=2.0)$.
 Each legend is a time interval $[T_{\min}^i,T_{\max}^i]$, where $T_{\min}^i$ and $T_{\max}^i$ are the minimum and maximum in the $i$th sorted data set $(i=1, \cdots, 13)$, respectively.}
\end{figure}

\begin{figure}
\includegraphics[height=.9\linewidth, angle=-90]{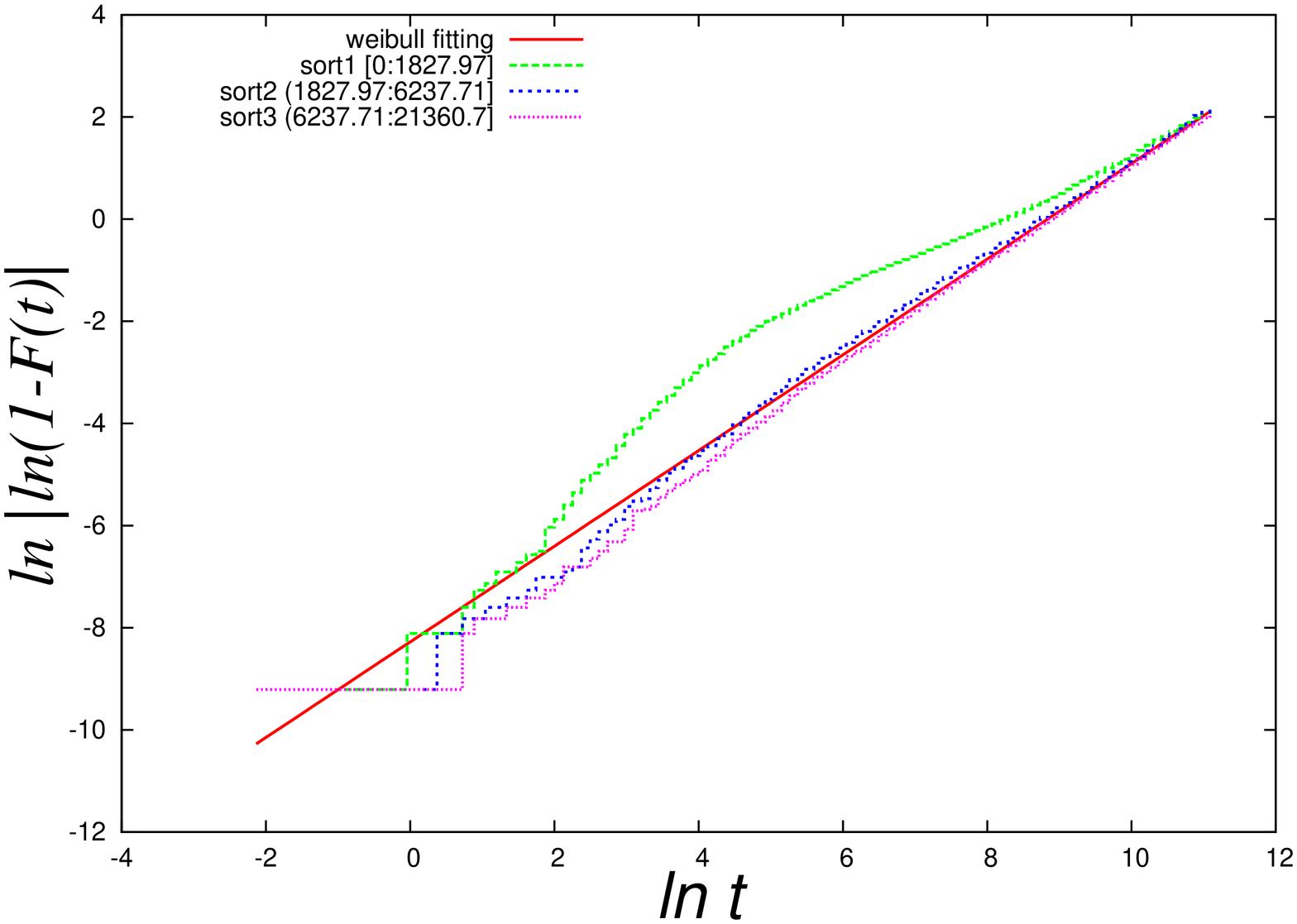}% Here is how to import EPS art
\caption{(color online). Weibull plot  of the distribution of
 interevent times of earthquakes in Japan $(M_c=3.0)$.
 %25${}^{\circ}$-50${}^{\circ}$N  and 125${}^{\circ}$-150${}^{\circ}$E  ($M_c=2.0$).
 Each legend is a time interval $[T_{\min}^i,T_{\max}^i]$, where $T_{\min}^i$ and $T_{\max}^i$ are the minimum and maximum in the $i$th sorted data set $(i=1, 2, 3)$, respectively.}
\end{figure}

\begin{table}[!t]
\caption{\label{tab:table1} The Weibull exponent $a$, the intensity of intermittency $\eta$, and the Lyapunov exponent $\lambda$.}
\begin{ruledtabular}
\begin{tabular}{ccccc}
 %&\multicolumn{2}{c}{$D_{4h}^1$}&\multicolumn{2}{c}{$D_{4h}^5$}\\
  $M_c$ & $a$ & $\eta$ &$\lambda$ & number of earthquakes \\ \hline
 2.0 & 0.977  & 0.023 &$4.7\times 10^{-3}$ & 130243 \\
 2.5 & 0.956  & 0.046 &$2.7\times 10^{-3}$ & 67912 \\
% 25${}^{\circ}$-50${}^{\circ}$N  and 125${}^{\circ}$-150${}^{\circ}$E &
 3.0 & 0.937  & 0.067 &$1.4\times 10^{-3}$ & 31857 \\
\end{tabular}
\end{ruledtabular}
\end{table}

\section{Conclusions}

Intermittency of renewal processes is studied by 
constructing one-dimensional piecewise linear maps from 
renewal processes. As a result, characterization of 
intermittent phenomena is extended to extremely heavy tail
 and stretched exponential relaxation phenomena, 
and we can also estimate the Lyapunov exponent of 
renewal processes, which measures the activity of events. \par
 Analyzing the occurrence of earthquakes, we found that the occurrence of earthquakes is not a renewal process, which is in agreement with \cite{Livina2005}.  However, interevent times are i.i.d. random variables when the previous interevent time is a relatively large. Moreover, the conditional probability distribution functions are characterized by the {\it Weibull distribution}, which means that the occurrence of earthquakes is {\it stationary very weak intermittency}. 
The intensity of  intermittency of earthquakes depends on the  threshold $M_c$. In particular, the 
intensity of intermittency increases monotonically with the threshold of 
magnitude, indicating the view that it is difficult to forecast the occurrence of large earthquakes.\par
To quantify intermittency in point processes, Bickel proposed a clear 
estimation of  intermittency using the correlation codimension 
\cite{Bickel1999}. Although we assume that  interevent times are 
i.i.d. random variables, which does not always hold  in point processes, 
we can study  underlying dynamics of intermittent phenomena with the aid of
  one-dimensional maps.  If we assume the 
integrate-and-fire model \cite{Sauer1994}, the orbit $x_n$ in dynamical systems is considered to be 
 the integrated value of a signal that is behind intermittent phenomena. In the occurrence of earthquakes, the integrated value would be the accumulated  energy on the crust.
\par

\begin{acknowledgments}
T.A. would like to thank N. Weissburg for fruitful discussions. 
We thank the JMA for allowing us to use the earthquake data.
This research was financially supported by the Sasakawa Scientific Research Grant from The Japan Science Society.
\end{acknowledgments}

%\newpage %Just because of unusual number of tables stacked at end


\bibliography{akimoto}

\begin{thebibliography}{26}
\bibitem[{FN2()}]{FN2}
\bibinfo{note}{In dynamical systems, the divergence of the mean interevent time
  means that the invariant measure can not be normalized, i.e., the infinite
  measure.}
\bibitem[{FN3()}]{FN3}
\bibinfo{note}{A positive function $L$ is slowly varying at $0$ iff
  ${\displaystyle \frac{L(tx)}{L(t)} \rightarrow 1}$, as $t\rightarrow 0$.}
\end{thebibliography}

\end{document}